\documentclass{llncs}
\usepackage{graphicx}
\usepackage{booktabs}
\usepackage{url}
\usepackage[font=small,compatibility=false,tableposition=top]{caption}
\usepackage{floatrow}
\usepackage[pdfborder={0 0 0}]{hyperref}
\usepackage{hyperref}
\usepackage{subfig}
\usepackage{xspace}
\usepackage{amssymb}
\usepackage{amsmath}
\usepackage{gensymb}
\usepackage[mathcal]{euscript}
\usepackage[nolist,nohyperlinks]{acronym}
\usepackage[ruled,vlined,linesnumbered]{algorithm2e}
\usepackage{todonotes}
\usepackage{multirow}
\usepackage{sidecap}
\usepackage{hyphenat}
\usepackage{etex}   %
\usepackage{tikz}
\usepackage{pdfpages}
\usepackage{framed}
\usepackage{textcomp}
\usepackage{wrapfig}

\spnewtheorem*{myproblem}{Problem}{\bfseries}{\rmfamily}

\DeclareRobustCommand{\romega}{\omega}
\DeclareRobustCommand{\repsilon}{\epsilon}
\newcommand{\wrev}{\ensuremath{\omega_{\mathrm{rev}}}\xspace}
\newcommand{\wlen}{\ensuremath{\omega_{\mathrm{len}}}\xspace}
\newcommand{\wwglay}[2]{$#1\text{-}#2$-\acs{glp}\xspace}

\newcommand{\eaga}{\acs{eaga}\xspace}

\newcommand{\bk}{BK\xspace}
\newcommand{\linseg}{LS\xspace}
\newcommand{\poly}{Poly\xspace}
\newcommand{\orth}{Orth\xspace}

\newcommand{\npc}{NP-complete\xspace}

\newcommand{\nph}{NP-hard\xspace}

\newcommand{\tsuc}{\textit{\scriptsize topSuc}}
\newcommand{\tpre}{\textit{\scriptsize topPre}}
\newcommand{\bsuc}{\textit{\scriptsize botSuc}}
\newcommand{\bpre}{\textit{\scriptsize botPre}}
\newcommand{\tadj}{\textit{\scriptsize topAdj}}
\newcommand{\badj}{\textit{\scriptsize botAdj}}

\newcommand{\pc}{{\footnotesize\,\%}\xspace}
\newcommand{\ta}{\kern0.2em}
\newcommand{\minuseq}{\raisebox{.4ex}{\tiny\ensuremath{%
-}}\nolinebreak\hspace{.05em}\raisebox{.4ex}{\tiny\ensuremath{=}}}
\newcommand{\incmm}{\raisebox{.4ex}{\tiny\ensuremath{%
-}}\nolinebreak\hspace{.05em}\raisebox{.4ex}{\tiny\ensuremath{-}}}
\newcommand{\incpp}{\raisebox{.4ex}{\tiny\ensuremath{%
+}}\nolinebreak\raisebox{.4ex}{\tiny\ensuremath{+}}}

\SetArgSty{textrm}

\SetCommentSty{commentsty}
\SetKw{Each}{each}
\ifdefined\SetAlCapSty
  \SetAlCapSty{textbf}
\fi
\ifdefined\SetAlgoCaptionSeparator
  \SetAlgoCaptionSeparator{.}
\fi
\ifdefined\DontPrintSemicolon
  \DontPrintSemicolon
\fi

\ifdefined\algorithmautorefname
  \renewcommand{\algorithmautorefname}{Alg.\@}
\else
  
\fi

\newcommand{\ie}{{i.\,e.\@}\xspace}
\newcommand{\eg}{{e.\,g.\@}\xspace}
\newcommand{\etal}{{et al.\@}\xspace}
\newcommand{\cf}{{cf.\@}\xspace}

\newcommand{\wrt}{{w.\,r.\,t.\@}\xspace}

\begin{acronym}
  \acro{aca}[ACA]{Adaptive Constrained Alignment}
  \acro{ascet}[ASCET]{Advanced Simulation and Control Engineering Tool}
  \acro{atl}[ATL]{Atlas Transformation Language}
  \acro{bfs}[BFS]{breadth first search}
  \acro{cau}[CAU]{Christian-Albrechts-Universit\"at zu Kiel}
  \acro{chess}[CHESS]{Center for Hybrid and Embedded Software Systems}
  \acro{codaflow}[CoDaFlow]{Constrained Data Flow}
  \acro{cola}[CoLa]{Constrained Layout}
  \acro{dag}[DAG]{directed acyclic graph}
  \acro{dfd}[DFD]{Data Flow Diagram}
  \acro{dfs}[DFS]{depth first search}
  \acro{dsl}[DSL]{Domain Specific Language}
  \acro{dsml}[DSML]{Domain Specific Modeling Language}
  \acro{ecu}[ECU]{electronic control unit}
  \acro{eecs}[EECS]{Electrical Engineering and Computer Sciences}
  \acro{ehandbook}[EHANDBOOK]{}
  \acro{emf}[EMF]{Eclipse Modeling Framework}
  \acro{etas}[ETAS]{Engineering Tools, Application and Services}
  \acro{fsm}[FSM]{finite state machine}
  \acro{gd}[GD]{graph drawing}
  \acro{gef}[GEF]{Graphical Editing Framework}
  \acro{glmm}[GLMM]{graph layout meta model}
  \acro{gmf}[GMF]{Graphical Modeling Framework}
  \acro{graphml}[GraphML]{Graph Markup Language}
  \acro{gui}[GUI]{Graphical User Interface}
  \acro{gxl}[GXL]{Graph eXchange Language}
  \acro{hmi}[HMI]{human-machine interface}
  \acro{ide}[IDE]{integrated development environment}
  \acro{ieee}[IEEE]{Institute of Electrical and Electronics Engineers}
  \acro{ilog}[ILOG]{Intelligence Logiciel}
  \acro{ilp}[ILP]{integer linear program}
  \acro{jni}[JNI]{Java Native Interface}
  \acro{json}[JSON]{Java Script Object Notation}
  \acro{jvm}[JVM]{Java Virtual Machine}
  \acro{lncs}[LNCS]{Lecture Notes in Computer Science}
  \acro{m2m}[M2M]{Model to Model}
  \acro{m2t}[M2T]{Model to Text}
  \acro{mde}[MDE]{model-driven engineering}
  \acro{mdsd}[MDSD]{model-driven software development}
  \acro{mic}[MIC]{model-integrated computing}
  \acro{moc}[MoC]{model of computation}
  \acro{mof}[MOF]{Meta Object Facility}
  \acro{mvc}[MVC]{model-view-controller}
  \acro{kaom}[KAOM]{\acs{kieler} Actor Oriented Modeling}
  \acro{kcss}[KCSS]{Kiel Computer Science Series}
  \acro{keg}[KEG]{\acs{kieler} Editor for Graphs}
  \acro{kiel}[KIEL]{Kiel Integrated Environment for Layout}
  \acro{kieler}[KIELER]{Kiel Integrated Environment for Layout Eclipse Rich Client}
  \acro{kiem}[KIEM]{\acs{kieler} Execution Manager}
  \acro{kiml}[KIML]{\acs{kieler} Infrastructure for Meta Layout}
  \acro{klay}[KLay]{\acs{kieler} Layouters}
  \acro{klaylay}[KLay Layered]{\acs{klay}\allowbreak Layered}
  \acro{klodd}[KLoDD]{\acs{kieler} Layout of Dataflow Diagrams}
  \acro{ocl}[OCL]{Object Constraint Language}
  \acro{ogdf}[OGDF]{Open Graph Drawing Framework}
  \acro{omg}[OMG]{Object Management Group}
  \acro{qvt}[QVT]{Query-View-Transformations}
  \acro{rca}[RCA]{Rich Client Application}
  \acro{rcp}[RCP]{Rich Client Platform}
  \acro{sbgn}[SBGN]{Systems Biology Graphical Notation}
  \acro{scade}[SCADE]{Safety Critical Application Development Environment}
  \acro{ssm}[SSM]{Safe State Machines}
  \acro{sud}[SUD]{system under development}
  \acro{tmf}[TMF]{Textual Modeling Framework}
  \acro{tsm}[TSM]{topology-shape-metrics}
  \acro{ui}[UI]{user interface}
  \acro{uml}[UML]{Unified Modeling Language}
  \acro{upl}[UPL]{Upward Planarization Layout}
  \acro{vlsi}[VLSI]{very large scale integration}
  \acro{wysiwyg}[WYSIWYG]{What-You-See-Is-What-You-Get}
  \acro{xgmml}[XGMML]{Extensible Graph Markup and Modeling Language}
  \acro{xml}[XML]{Extensible Markup Language}
\end{acronym}

\acrodef{fas}[FAS]{Feedback Arc Set}
\acrodef{fasp}[FASP]{Feedback Arc Set Problem}
\acrodef{mfas}[MFAS]{Minimum Feedback Arc Set}
\acrodef{scp}[SCP]{Sum Coloring Problem}
\acrodef{bscp}[BSCP]{Bandwidth Sum Coloring Problem}
\acrodef{dlp}[DLP]{Directed Layering Problem}
\acrodef{glp}[GLP]{Generalized Layering Problem}
\acrodef{glpip}[GLP-IP]{Generalized Layering Problem Integer Program}
\acrodef{glph}[GLP-H]{Generalized Layering Problem Heuristic}
\acrodef{lap}[LAP]{Linear Arrangement Problem}
\acrodef{olap}[OLAP]{Optimal Linear Arrangement Problem}

\acrodef{eaga}[EaGa]{Eades and Gansner}
\acrodef{optga}[OptGa]{Optimal Cycle Breaking and Gansner}
\acrodef{glay}[GLay]{General Layering}
\acrodef{dlay}[DLay]{Directed Layering}
\acrodef{wwglay}[$\romega_{\mathrm{len}}$-$\romega_{\mathrm{rev}}$-GLay]{General Layering with weights}
\acrodef{eglay}[$\repsilon$-GLay]{}

\begin{document}
\mainmatter

\title{A Generalization of the \\Directed Graph Layering Problem}

\author{Ulf R\"uegg\inst{1} \and Thorsten Ehlers\inst{1}\and \\
  Miro Sp{\"o}nemann\inst{2} \and Reinhard von Hanxleden\inst{1}}

\institute{
  Dept.\@ of Computer Science, Kiel University, Kiel, Germany\\
  \email{\{uru,the,rvh\}@informatik.uni-kiel.de}
  \and
  TypeFox GmbH, Kiel, Germany\\
  \email{miro.spoenemann@typefox.io}
}

\maketitle

\setcounter{footnote}{0}

\begin{abstract}
The \acf{dlp} solves a step of the
widely used layer-based approach to automatically
draw directed acyclic graphs. To cater for cyclic graphs,
usually a preprocessing step is used that solves
the \acf{fasp} to make the graph acyclic before
a layering is determined.

Here we present the \acf{glp},
which solves the combination of \acs{dlp} and \acs{fasp} simultaneously,
allowing general graphs as input.
We present an integer programming model
and a heuristic to solve the \npc \acs{glp} and perform
thorough evaluations on different sets of graphs and with
different implementations for the steps of the layer-based approach.

We observe that \acs{glp} reduces the number of
dummy nodes significantly, can produce more compact drawings, and
improves on graphs where \acs{dlp} yields poor aspect ratios.


\end{abstract}

\section{Introduction}
\label{sec:introduction}

\begin{wrapfigure}[28]{R}{.5\textwidth}
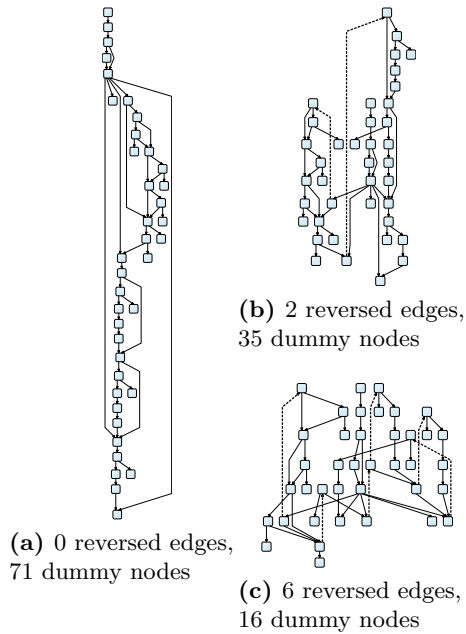

    \centering
    \begin{minipage}[t]{0.48\textwidth}
      \subfloat[][0 reversed edges,\\71 dummy nodes]{
        \label{fig:first_example_1}
        \begin{minipage}{\textwidth}
          \centering
          \includegraphics[scale=0.15]{{{images/g.39.29_eaga_r0_d71}}}
        \end{minipage}
      }
    \end{minipage}
    \begin{minipage}[t]{0.48\textwidth}
      \subfloat[][2 reversed edges,\\35 dummy nodes]{
        \begin{minipage}{\textwidth}
          \centering
          \includegraphics[scale=0.15]{{{images/g.39.29_wd1_wr10_r2_d35}}}
        \end{minipage}
      }\\
      \subfloat[][6 reversed edges,\\16 dummy nodes]{
        \begin{minipage}{\textwidth}
          \centering
          \includegraphics[scale=0.15]{{{images/g.39.29_wd1_wr2_r6_d16}}}
        \end{minipage}
      }
    \end{minipage}
  \caption{
    Different drawings of the \textsf{g.39.29} graph from the
    North graphs collection~\cite{DiBattistaGLP+97}.
    (a)~is drawn with known methods~\cite{GansnerKNV93},
    (b)~and (c) are results of the methods presented here.
    Backward edges are drawn bold and dashed.
  }%
  \label{fig:first_example}%
\end{wrapfigure}

The layer-based approach is a well-established and widely used
method to automatically draw directed graphs.
It is based on the idea to assign nodes to subsequent \emph{layers}
that show the inherent direction of the graph,
see \autoref{fig:first_example_1} for an example.
The approach was introduced by Sugiyama \etal~\cite{SugiyamaTT81} and
remains a subject of ongoing research.

Given a directed graph, the layer-based approach
was originally defined for acyclic graphs as a pipeline of three phases. However,
two additional phases are necessary to allow practical usage,
which are marked with asterisks:
\begin{enumerate}
  \item \emph{Cycle removal*:}
      Eliminate all cycles by reversing a preferably small subset of the graph's edges.
      This phase adds support for cyclic graphs as input.
  \item \emph{Layer assignment:}
      Assign all nodes to numbered \emph{layers} such that edges point from
      layers of lower index to layers of higher index. Edges
      connecting nodes that are not on consecutive layers
      are split by so-called \emph{dummy nodes}.
  \item \emph{Crossing reduction:}
      Find an ordering of the nodes within each layer such that the
      number of crossings is minimized.
  \item \emph{Coordinate assignment:}
      Determine explicit node coordinates with the goal to
      minimize the distance of edge endpoints.
  \item \emph{Edge routing*:}
      Compute bend points for edges, \eg
      with an orthogonal style.
\end{enumerate}

While state-of-the-art methods produce drawings
that are often satisfying, there are graph instances
where the results show bad \emph{compactness} and
unfavorable \emph{aspect ratio}~\cite{GutwengervHM+14}.
In particular, the number of layers is bound
from below by the longest path
of the input graph after the first phase. When placing the
layers vertically one above the other, this
affects the height of the drawing, see
\autoref{fig:first_example_1}.
Following these observations, we present new methods
to overcome current limitations.

\paragraph{Contributions.}

The focus of this paper is on the first two phases stated above.
They determine the initial topology of the drawing
and thus directly impact
the compactness and the aspect ratio
of the drawing.

\begin{wrapfigure}[26]{r}{.5\textwidth}
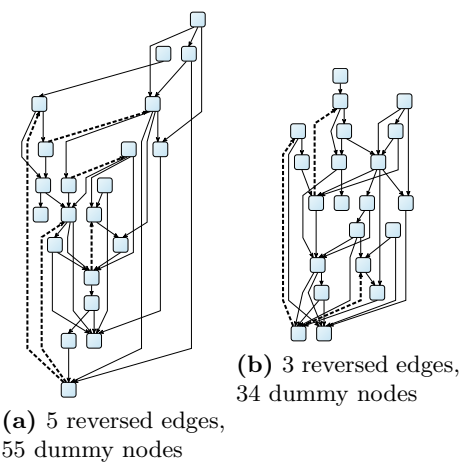

  \centering
    \centering
    \subfloat[5 reversed edges, 55 dummy nodes]{
      \begin{minipage}{.45\textwidth}
        \centering
        \includegraphics[scale=0.24]{{{images/random25_13_eaga_r5_d55}}}
      \end{minipage}
    }\hfill
    \subfloat[3 reversed edges, 34 dummy nodes]{
      \begin{minipage}{.45\textwidth}
        \centering
        \includegraphics[scale=0.24]{{{images/random25_13_1-30glay_r3_d34}}}
      \end{minipage}
    }%
  \caption{A graph drawn with (a) \eaga (known methods as
           described in \autoref{sec:related_work}) and (b) \wwglay{1}{30} (this work).
           This example illustrates that \acs{glpip} can perform better
           in both metrics: reversed edges (dashed) and dummy nodes.
           }
  \label{fig:example_glay}
\end{wrapfigure}

We introduce a new layer assignment method
which is able to handle cyclic graphs
and to consider compactness properties for selecting an edge reversal set.
Specifically, 1)~it can overcome the previously
mentioned lower bound on
the number of layers arising from the longest path
of a graph, 2)~it can be flexibly configured to
either favor elongated or narrow drawings, thus
improving on aspect ratio, and 3)~compared to
previous methods it is able to reduce
both the number of dummy nodes and
reversed edges
for certain graphs.
See \autoref{fig:first_example} and
\autoref{fig:example_glay} for examples.

We discuss how to solve
the new method
to optimality using an
integer programming model
as well as heuristically,
and evaluate both.

\paragraph{Outline.}
The next section presents related work.
We introduce problems and definitions in \autoref{sec:preliminaries},
and present methods to solve the newly
introduced problems in \autoref{sec:glay} and \autoref{sec:heuristic}.
\autoref{sec:evaluations} discusses
thorough evaluations before we conclude in \autoref{sec:conclusion}.

\section{Related Work}
\label{sec:related_work}

The cycle removal phase targets the \npc \acf{fasp}.
Several approaches
have been proposed to solve \acs{fasp}
either to optimality or heuristically~\cite{HealyN13}.
In the context of layered graph drawing,
reversing
a minimal number of edges does not necessarily
yield the best results,
and application-inherent information
might make certain edges better candidates
to be reversed~\cite{GansnerKNV93}.
Moreover, the decision which edges to
reverse in order to make a graph acyclic
has a big impact on the results of the
subsequent layering phase.
Nevertheless the two phases are
executed separately until today.

To solve the second phase, \ie the layer assignment
problem, several approaches with
different optimization goals have emerged.
Eades and Sugyiama employ a
longest path layering, which requires linear
time, and the resulting number
of layers equals the number of nodes of the graph's
longest path~\cite{EadesS90}.
Gansner \etal solve the layering
phase by minimizing the sum of the edge lengths
regarding the number of necessary dummy nodes~\cite{GansnerKNV93}.
They show that the problem is solvable in polynomial time and
present a network simplex algorithm which in turn
is not proven to be polynomial, although
it runs fast in practice.
This approach was found to inherently produce
compact drawings and performed best in comparison
to other layering approaches~\cite{HealyN02a}.

Healy and Nikolov tackle the problem of finding
a layering subject to bounds on the number of
layers and the maximum number of nodes in any layer
with consideration of dummy nodes
using an integer linear programming approach~\cite{HealyN02a}.
The problem is \nph, even without
considering dummy nodes.
In a subsequent
paper they present a
branch-and-cut algorithm to
solve the problem faster and for larger
graph instances~\cite{HealyN02b}.
Later, Nikolov~\etal
propose and evaluate
several heuristics to find a layering with
a restricted number of nodes in each layer~\cite{NikolovTB05}.
Nachmanson \etal present
an iterative algorithm to produce drawings with
an aspect ratio close to a previously specified
value \cite{NachmansonRL08}.

All of the previously mentioned layering methods have
two major drawbacks.
1)~They require the input graph to be
acyclic upfront, and
2)~they are bound to a minimum number
of layers equal to the longest path of the graph.
In particular this means that the bound
on the number of layers in the methods of Nikolov \etal
cannot be smaller than the longest path.

In the context of force-directed layout,
Dwyer and Koren presented a method
that can incorporate constraints
enforcing all directed edges
to point in the same direction~\cite{DwyerK05}.
They explored the possibility to relax some
of the constraints,
\ie let some of the edges point backwards,
and found that this
improves the readability of the drawing.
In particular, it reduced the number of edge crossings.

\section{Definitions and Problem Classification}
\label{sec:preliminaries}

Let $G=(V,E)$ denote a graph with a set of nodes $V$
and a set of edges $E$. We write
an edge between nodes $u$ and $v$ as $(u,v)$
if we care about direction, as $\{u,v\}$ otherwise.
A~\emph{layering} of a directed graph $G$ is a
mapping $L: V \rightarrow \mathbb{N}$.
A~layering $L$ is \emph{valid}
if $\forall (u,v) \in E$: $L(v) - L(u) \geq 1$.

\begin{myproblem}[\textbf{Directed Layering (DLP)}]
  Let $G = (V,E)$ be an acyclic directed graph.
  The problem is to find a minimum $k$ and a
  valid layering $L$ such that
  $\sum_{(v,w) \in E} (L(w) - L(v)) = k$.
\end{myproblem}

\noindent
As mentioned in \autoref{sec:related_work},
\acs{dlp} was originally introduced by
Gansner~\etal~\cite{GansnerKNV93}.
We extend the idea of a layering for
directed acyclic graphs to general graphs, \ie
graphs that are either directed or undirected and
that can possibly be cyclic.
Undirected graphs can be handled by assigning
an arbitrary direction to each edge,
thus converting it into a directed one,
and by hardly penalizing reversed edges.
We call a layering $L$ of a general graph $G$ \emph{feasible}
if $\forall \{u,v\} \in E: |L(u) - L(v)| \geq 1$.

\begin{myproblem}[\textbf{Generalized Layering (GLP)}]
  Let $G = (V,E)$ be a possibly cyclic directed graph and
  let $\wlen, \wrev \in \mathbb{N}$ be weighting constants.
  The problem is to find a minimum $k$ and a feasible layering $L$ such that
  $$\wlen\left( \sum_{(v,w) \in E} \left|L(w) - L(v)\right| \right)
   + \wrev\left|\left\{ (v,w) \in E: L(v) > L(w) \right\}\right| \ = \ k \enspace.$$
\end{myproblem}

Intuitively, the left part of the sum represents the overall
edge length (\ie the number of dummy nodes) and the right part represents
the number of reversed edges (\ie the \acs{fas}).
After reversing all edges in this \acs{fas},
the feasible layering becomes a valid layering.
Compared to the standard cycle removal phase combined with \acs{dlp},
the generalized layering problem allows more flexible
decisions on which edges to reverse.
Also note that \acs{glp} with $\wlen = 1, \wrev = \infty$
is equivalent to \acs{dlp} for acyclic input graphs and that
while \acs{dlp} is solvable in polynomial time, both
parts of \acs{glp} are \npc \cite{RueeggESvH15}.
\section{The IP Approach}
\label{sec:glay}
In the following, we describe how to solve \acs{glp} using integer programming.
The rough idea of this model is to assign integer values
to the nodes of the given graph that represents the layer
in which a node is to be placed.

\smallskip
\noindent\textit{Input and parameters.}
Let $G=(V,E)$ be a graph with node set $V = \{1,\dots,n\}$.
Let $e$ be the
adjacency matrix, \ie $e(u,v) = 1$ if $(u,v) \in E$ and $e(u,v) = 0$ otherwise.
$\wlen$ and $\wrev$ are weighting constants.

\smallskip
\noindent\textit{Integer decision variables.}
$l(v)$ takes a value in $\{1,\dots,n\}$ indicating that
node $v$ is placed in layer $l(v)$, for all $v \in V$.

\smallskip
\noindent\textit{Boolean decision variables.}
$r(u,v) = 1$ if and only if edge $e = (u,v) \in E$ and $e$ is reversed, \ie
$l(u) > l(v)$, for all $u,v \in V$. Otherwise, $r(u,v) = 0$.
\begin{eqnarray*}
  \small
 \text{Minimize} \qquad \wlen \sum_{(u,v) \in E} |l(u) - l(v)|
                +\ \wrev \sum_{(u,v) \in E} r(u,v) \text{.}
\end{eqnarray*}

The sums represent the edge lengths, \ie the number of dummy nodes, and
the number of reversed edges, respectively.
Constraints are defined as follows:
\begin{eqnarray}
  \small
   1 \ \leq\ l(v) \ \leq\  n \quad && \forall v\in V  \label{eqn:first} \\
   |l(u) - l(v)| \ \geq\  1 \quad && \forall (u,v) \in E\quad \label{eqn:second} \\
   n \cdot r(u,v) + l(v) \ \geq\  l(u) + 1 \quad && \forall (u,v) \in E \label{eqn:third}
\end{eqnarray}
Constraint~(\ref{eqn:first}) restricts the range of possible layers.
(\ref{eqn:second})~ensures that the resulting layering is feasible.
(\ref{eqn:third})~binds the decision variables in $r$ to the layering,
\ie because $r$ is part of the objective, and $\wrev > 0$,
$r(u,v)$ gets assigned 0 unless $l(v) < l(u)$, for all $(u,v) \in E$.

\paragraph{Variations.}
\vspace*{-8pt}
The model can easily be extended to restrict the number of
layers by replacing the $n$ in constraint (1) by
a desired bound $b \leq n$.

The edge matrix can be extended to
contain a weight $w_{u,v}$ for each edge $(u,v) \in E$.
This can be helpful if further semantic information is available,
\ie about feedback edges that lend themselves well to be reversed.

\section{The Heuristic Approach}
\label{sec:heuristic}

Interactive modeling tools providing automatic layout facilities
require execution times significantly shorter than one second.
As the IP formulation discussed in the previous section rarely meets
this requirement, we present a heuristic
to solve \acs{glp}.
It proceeds as follows.
1)~Leaf nodes are removed iteratively, since it is trivial
to place them with minimum edge length and desired edge direction.
Note that therefore the heuristic is not yet able to improve on trees that yield
a poor compactness. We leave this for future research.
2)~For the (possibly cyclic) input graph an initial feasible layering
is constructed which is used to deduce edge directions
yielding an acyclic graph.
3)~Using the network simplex
method presented by Gansner \etal \cite{GansnerKNV93},
a solution with minimal edge length is created.
4)~We execute a greedy improvement procedure
after which we again deduce edge directions and re-attach the leaves.
5)~We apply the
network simplex algorithm a second time to
get a valid layering with minimal edge lengths for
the next steps of the layer-based approach.
In the following we will discuss steps 2 and 4 in further detail.

\paragraph{Step 2: Layering Construction.}
To construct an initial feasible solution we follow
an idea that was first presented by McAllister as
part of a greedy heuristic for the \acf{lap}~\cite{McAllister99}
and later extended by Pantrigo \etal~\cite{PantrigoMDP12}.

Nodes are assigned to distinct indexes,
where as a start, a node is selected randomly, assigned
to the first index, and added to a set of assigned nodes.
Based on the set of assigned nodes
a candidate list is formed, and the most promising node
is assigned to the next index.
As decision criterion we use the difference between
the number of edges incident to unassigned nodes
and the number of edges incident to assigned nodes.
This procedure is repeated
until all nodes are assigned to distinct indices
(see \autoref{alg:constructLayering}).

In contrast to McAllister, for \acs{glp} we allow nodes
to be added to either side of the set of assigned nodes,
and decide the side based on the number
of reversed edges that would emerge from
placing a certain node on that side.
For this we use a decreasing left index variable
and an increasing right index variable.

\begin{algorithm}[tb]
  \scriptsize
  \KwIn{directed graph $G=(V,E)$}
  \KwData{
    Sets $U$, $C$. For all $v \in V$ $score[v], incAs[v], outAs[v]$\;
    $lIndex \leftarrow -1$, $rIndex \leftarrow 0$\;
  }
  \KwOut{index[v]: feasible layering of G}
  \vspace{.5em}
  \For{$v \in V$} {
    $score[v] \leftarrow |\{ w \enspace | \enspace \{v,w\} \in E \}|$; $incAs[v] \leftarrow 0$; $outAs[v] \leftarrow 0$\;
    add $v$ to $U$
  }
  remove random $v$ from $U$\;
  $c \leftarrow v$\;
  \While{$U$ not empty}{
    \eIf{$incAs[c] < outAs[c]$}{ $index[c] \leftarrow lIndex$\incmm }{ $index[c] \leftarrow rIndex$\incpp }
    remove $c$ from $U$ and $C$; $cScore \leftarrow \infty$\;
    \For{$v \in \{ w \enspace | \enspace \{c, w\} \in E \wedge w \in U \}$} {
      add $v$ to $C$; $score[v]$\incmm\;
      \leIf{$(c,v) \in E$}{$incAs[v]$\incpp}{$outAs[v]$\incpp}
    }
    \For{$v \in C$} {
      \lIf{ $score[v] < cScore$ }{
        $cScore \leftarrow score[v]$; $c \leftarrow v$
      }
    }
  }
  \caption{constructLayering}
  \label{alg:constructLayering}
\end{algorithm}

\paragraph{Step 4: Layering Improvement.}
At this point a feasible layering with a minimum number
of dummy nodes \wrt the chosen \acs{fas} is given since we execute
the network simplex method of Gansner \etal beforehand.
Thus we can only improve on
the number of reversed edges. We determine possible
\emph{moves} and decide whether to take the move
based on a \emph{profit} value.
Let a graph $G=(V,E)$ and a feasible layering $L$ be given.
For ease of presentation, we define the following notions.
An example:
For a node $v$,
$\tsuc$ are the nodes connected to $v$
via an outgoing edge of $v$
and are currently assigned to a layer
with lower index than $v$'s index.
Intuitively, $\tsuc$ (just as $\bpre$) are nodes connected
by an edge pointing into the ``wrong'' direction.
{\small
\begin{align*}
	&v.\tsuc = \{ w : (v,w) \in E \wedge L(v) > L(w) \}
	&&v.\bsuc = \{ w : (v,w) \in E \wedge L(v) < L(w) \} \\
	&v.\tpre = \{ w : (w,v) \in E \wedge L(w) < L(v) \}
	&&v.\bpre = \{ w : (w,v) \in E \wedge L(w) > L(v) \} \\
	&v.\tadj = v.\tsuc \cup v.\tpre
	&&v.\badj = v.\bsuc \cup v.\bpre
\end{align*}
}

For all these functions we define suffixes that allow to query
for a certain set of nodes before or after a certain index. For instance, for
all top successors of $v$ before index $i$ we write
$v.\textit{\scriptsize topSucBefore(i)} = \{ w : w \in v.\tsuc \wedge L(w) < i \}$.

Let $move : V \mapsto \mathbb{N}$ denote a function
assigning to each node a natural value.
The function describes whether it is possible
to move a node without violating the layering's feasibility
as well as how far the node should be moved.
For instance,
let for a node $v$ $\tpre$ be empty but
$\tsuc$ be not empty.
Thus, we can move $v$ to an arbitrary layer with
lower index than $L(v)$. A good choice
would be one layer before any
of $v$'s $\tsuc$ since this would alter
the connected edges to point downwards.
{\small
\begin{align*}
	move(v) = \left\{ \begin{array}{l l}
		    0 & \quad \text{if } v.\tsuc = \emptyset, \\
		    L(v) - \min(\{L(w) : w \in v.\tsuc \}) + 1 & \quad \text{if } v.\tpre = \emptyset, \\
		    L(v) - \max(\{L(w) : w \in v.\tpre \}) -  1 & \quad \text{otherwise.}
		  \end{array} \right.
\end{align*}
}

Let
$\mbox{\itshape profit} : V \times \mathbb{N} \times \mathbb{N} \mapsto \mathbb{Z}$ denote a function
assigning a quality score to each node $v$ if it were
moved by $m \in \mathbb{N}$ to a different layer $x$,
\ie if it is worth to increase some edges' lengths for
a subset of them to point downwards.
Note that we reuse \wlen and \wrev here but do not expect them
to have an impact as strong as for the IP. For the rest of
the paper we fix them to 1 and 5.
{\small
\begin{align*}
	\mbox{\itshape profit}(v, m, x) = \left\{ \begin{array}{l l}
		    0 & \quad \text{if } m \leq 1, \\
		    \wlen (m | v.\textit{\scriptsize topAdjBefore(x)} | - m | v.\badj | ) & \quad \\
		    + \enspace \wrev | v.\textit{\scriptsize topSucAfter(x)} | & \quad \text{otherwise.}
		  \end{array} \right.
\end{align*}
}

As seen in \autoref{alg:improveLayering},
the $move$ and $\mbox{\itshape profit}$ functions are determined
initially for a given feasible layering.
A queue, sorted based on profit values,
is then used to successively perform moves that yield
a profit.
After a move of node $n$, both functions can be
updated for all nodes in the adjacency of $n$.

\begin{algorithm}[tb]
  \scriptsize
  \KwIn{feasible layering of $G=(V,E)$ in $index[v]$}
  \KwData{
    priority queue PQ\;
    For all $v \in V$ $move[v]$, $\mbox{\textit{profit}}[v]$
  }
  \KwOut{index[v]: feasible layering of G}
  \vspace{.5em}
  \For{$v \in V$} {
    $move[v] \leftarrow move(v)$\;
    $\mbox{\textit{profit}}[v] \leftarrow \mbox{\textit{profit}}(v,\ move[v],\ index[v] - move[v])$\;
    \lIf{ $\mbox{\textit{profit}}[v] > 0$ } { enqueue $v$ to $PQ$ }
  }
  \While{$PQ$ not empty} {
    $v \leftarrow$ dequeue $PQ$\;
    $index[v]$ \minuseq $\enspace move[v]$\;
    \For{$w \in \{ w \enspace | \enspace \{v,w\} \in E \}$} {
      update $move[w]$ and $\mbox{\textit{profit}}[w]$\;
      \leIf{ $\mbox{\textit{profit}}[w] > 0$ }{ enqueue $w$ to $PQ$ } { possibly dequeue $w$ from $PQ$}
    }
  }
  \caption{improveLayering}
  \label{alg:improveLayering}
\end{algorithm}

\paragraph{Time Complexity.}
\vspace{-8pt}
Removing leaf nodes requires linear time, $O(|V|+|E|)$.
\autoref{alg:constructLayering} is
quadratic in the number of nodes, $O(|V|^2)$. The while loop
has to assign an index to every node and the two inner
for loops are, for a complete graph, iterated $\frac{|V|}{2}$
times on average. Determining the next candidate (lines 15--16)
could be accelerated using dedicated data structures.
The improvement step strongly depends on the
input graph.
The network simplex method runs reportedly fast in practice~\cite{GansnerKNV93},
although it has not been proven to be polynomial.
Our evaluations showed
that the heuristic's overall execution time is clearly
dominated by the network simplex method (\cf \autoref{sec:execution_times}).

\section{Evaluation}
\label{sec:evaluations}

In this section we evaluate three points: 1)~the general feasibility
of \acs{glp} to improve the compactness of drawings,
2)~the quality of metric estimations for area and aspect ratio,
and 3)~the performance of the presented IP and heuristic.
Our main metrics of interest here are height, area, and
aspect ratio, as defined in an earlier paper~\cite{GutwengervHM+14}.
Remember that the layer-based approach is defined as a pipeline
of several independent steps.
After the layering phase,
which is the focus of our research here,
these latter two metrics can only be estimated
using the number of
dummy nodes, the number of layers, and the maximal
number of nodes in a layer.
Results can be seen in \autoref{tab:common_results}
and \autoref{tab:metric_results}, which we will discuss
in more detail in the remainder of this section.

\paragraph{Obtaining a Final Drawing.}
To collect all metrics we desire,
we have to create a final drawing of a graph.
Over time numerous strategies
have been presented for each step
of the layer-based approach,
we thus present several alternatives.
To break cycles we use a popular heuristic by Eades \etal~\cite{EadesLS93}.
To determine a layering we use our newly presented
approach \acs{glp} (both the IP method and heuristic,
denoted by \acs{glpip} and \acs{glph})
and alternatively
the network simplex method presented by Gansner
\etal~\cite{GansnerKNV93}.
We denote the combination of the cycle breaking of Eades \etal
and the layering of Gansner \etal as \eaga and consider it to be
an alternative to \acs{glp}.
Crossings between pairs of layers are minimized
using a layer sweep method in conjunction with the
barycenter heuristic, as originally proposed by
{Sugiyama~\etal~\cite{SugiyamaTT81}}.
We employ two different strategies
to determine fixed coordinates
for nodes within the layers.
First, we consider a method introduced by Buchheim \etal
that was extended by Brandes and K{\"o}pf~\cite{BuchheimJL01,BrandesK02},
which we denote as \bk.
Second, we use a method inspired by Sander~\cite{Sander96a} that
we call \linseg.
Edges are routed either using polylines (\poly)
or orthogonal segments (\orth).
The orthogonal router is based on the methods presented by
Sander~\cite{Sander04}.
Overall, this gives twelve setups of the algorithm: three layering methods,
two node placement algorithms, and two edge routing procedures.
In the following, let \mbox{\wlen-\wrev-\acs{glp}} denote the used weights.
If we do not further qualify \acs{glp}, we refer to the IP model.

\setlength{\tabcolsep}{.4em}
\begin{table}[t]
  \caption{
    Average values for different layering strategies employed to
    the test graphs.
    Different weights are used for \acs{glpip} as specified in the column head and final drawings
    were created using \bk and \poly.
    For $\text{\acs{glph}}^{*}$ no improvement was performed.
    A detailed version of these results is included in the appendix (\cf \ref{sec:detailed_results}).
  }
  \label{tab:common_results}
  \subfloat[Random graphs]{
    \label{tab:random_results}
    \centering
    \scriptsize
    \begin{tabular}{l r r r r r r r r}
      \toprule
                     & 1-10 & 1-20 & 1-30 & 1-40 & 1-50 & \eaga & \acs{glph} & $\text{\acs{glph}}^{*}$\\
      \midrule
      Reversed edges &  3.71 &  2.89 &  2.64 &  2.54 &  2.44 &  2.93 &  8.67 & 10.36 \\
      Dummy nodes    & 34.45 & 46.73 & 52.79 & 56.14 & 60.53 & 72.64 & 48.48 & 58.21 \\
      \midrule
      Height         & 843 & 943 & 980 & 1,004 & 1,025 & 1,084 & 930 & 1,027\\
      Area           & 631,737 & 672,717 & 691,216 & 700,385 & 708,361 & 737,159 & 656,070 & 720,798\\
      Aspect ratio   & 0.77 & 0.65 & 0.63 & 0.61 & 0.59 & 0.55 & 0.67 & 0.60\\
      \bottomrule
    \end{tabular}
  }\\
  \subfloat[North graphs]{
    \label{tab:highar_results}
    \centering
    \scriptsize
    \begin{tabular}{l r r r r r r r r}
      \toprule
                     & 1-10    & 1-20 & 1-30 & 1-40 & 1-50 & \eaga & GLP-H & $\text{GLP-H}^{*}$ \\
      \midrule
      Reversed edges & 2.74    & 1.47 & 1.02 & 0.72 & 0.56 & 0 & 7.07 & 8.55\\
      Dummy nodes    & 39.91   & 55.47 & 65.73 & 75.66 & 82.47 & 141.30 & 53.53 & 68.91 \\
      \midrule
      Height         & 1,068   & 1,224 & 1,334 & 1,409 & 1,469 & 1,727 & 1,137 & 1,216 \\
      Area           & 587,727 & 622,838 & 641,581 & 660,842 & 695,494 & 874,374 & 629,778 & 691,372 \\
      Aspect ratio   & 0.34    & 0.28 & 0.24 & 0.23 & 0.22 & 0.20 & 0.33 & 0.32 \\
      \bottomrule
   \end{tabular}
  }
\end{table}

\paragraph{Test Graphs.}
Our new approach is intended to improve the drawings
of graphs with a large %
height and relatively small width, hence
unfavorable aspect ratio.
Nevertheless, we also evaluate the generality of the approach using
a set of 160 randomly generated graphs with 17 to 60 nodes and an
average of $1.5$ edges per node.
The graphs were generated by creating a number of nodes,
assigning out-degrees to each node such that the sum of outgoing
edges is $1.5$ times the nodes, and finally creating
the outgoing edges with a randomly chosen target node.
Unconnected nodes were removed.
Second, we filtered the graph set provided by North%
\footnote{\url{http://www.graphdrawing.org/data/}} \cite{DiBattistaGLP+97}
based on
the aspect ratio and selected 146 graphs that have at least 20 nodes
and a drawing\footnote{Created using \bk and \poly.}
with an aspect ratio below $0.5$,
\ie are at least twice as high as wide.
We also removed plain paths, that is, pairs of nodes
connected by exactly
one edge, and trees. For these special cases \acs{glp} in its
current form would not change the resulting number of reversed edges
as all edges can be drawn with length $1$. This is also true
for any bipartite graph.
Note however that \acs{glp} can easily incorporate
a bound on the number of layers which can straightforwardly
be used to force more edges to be reversed,
resulting in a drawing with better aspect ratio.

\paragraph{General Feasibility of GLP.}
An exemplary result of the \acs{glp} approach
compared to \eaga can be seen in \autoref{fig:example_glay}.
For that specific drawing, \acs{glp} produces fewer reversed
edges, fewer dummy nodes, and less area (both in width and height).
For all tested setups the average effective height and
area (normalized by the number of nodes)
of \acs{glp} and the heuristic are smaller than \eaga's,
see \autoref{tab:metric_results}.
The average aspect ratios come closer to $1.0$.
For simplicity, in this paper we desire aspect ratios
closer to $1.0$. For a more detailed discussion
on this topic see Gutwenger~\etal~\cite{GutwengervHM+14}.

Furthermore, we found that by altering
the weights \wrev and \wlen a
trade-off between reversed edges and resulting dummy nodes
(and thus area and aspect ratio)
can be achieved, which can be seen in
Tab.\ref{tab:random_results}.

The results for the North graphs are similar.
Since the North graphs are acyclic,
the cycle breaking phase
is not required and current layering algorithms
cannot improve the height.
The \acs{glp} approach, however, can freely reverse edges
and hereby change the height and aspect ratio.
Results can be seen in Tab.~\ref{tab:highar_results}.
Clearly, \eaga has no reversed edges as all graphs are acyclic.
\wwglay{1}{10} starts with an average of $2.7$ reversed
edges and the value constantly decreases with an increased
weight on reversed edges.
The number of dummy nodes on the other hand constantly decreases
from $141.3$ for \eaga to $39.9$ for \wwglay{1}{10}.

The average height and average area of the final drawings decrease with
an increasing number of reversed edges. For \wwglay{1}{10}
the average height and area are
38.2\pc and 33.8\pc smaller than \eaga.
The aspect ratio changes from an average of 0.20 for \eaga to 0.34
for \wwglay{1}{10}.

The results show that for the selected graphs, for which
current methods cannot improve on height, the
weights of the new approach allow to find
a satisfying trade-off between reversed edges and dummy nodes.
Furthermore, the improvements in compactness
stem solely from the selection of weights, not from
an upper bound on the number of layers. Naturally,
such a bound can further improve the aspect ratio and height.

\setlength{\tabcolsep}{.1em}
\begin{table}[t]
  \centering
  \scriptsize
  \caption{
    Results for final drawings of the set of random graphs,
    when applying different layout strategies.
    For \acs{glpip} $\wlen = 1$ and $\wrev = 30$ were used.
    Area is normalized by a graph's node count.
    The most interesting comparisons are between columns
    where \eaga and \acs{glp} use the same strategies for the
    remaining steps.
    Detailed results can be found in the appendix (\cf \ref{sec:detailed_results}).
    }
  \label{tab:metric_results}
  \begin{tabular}{ l r r r r r r r r r r r r }
    \toprule
      {\tiny Edge routing} & \multicolumn{6}{c}{ Poly } & \multicolumn{6}{c}{ Orth } \\ \cmidrule(lr){2-7} \cmidrule(lr){8-13}%
      {\tiny Node coord.} & \multicolumn{3}{c}{ BK } & \multicolumn{3}{c}{ LS } & \multicolumn{3}{c}{ BK } & \multicolumn{3}{c}{ LS } \\ \cmidrule(lr){2-4}  \cmidrule(lr){5-7} \cmidrule(lr){8-10} \cmidrule(lr){11-13}%
      {\tiny Layering} & {\tiny\eaga} & {\tiny \acs{glpip}} & {\tiny\acs{glph}} & {\tiny\eaga} & {\tiny \acs{glpip}} & {\tiny\acs{glph}} & {\tiny\eaga} & {\tiny \acs{glpip}} & {\tiny\acs{glph}} & {\tiny\eaga} & {\tiny \acs{glpip}} & {\tiny\acs{glph}}\\
    \midrule
    Height  & 1,165 & 1,043 & 898 & 943 & 824 & 732 & 790 & 711 & 652 & 817 & 746 & 678 \\
    Area         & 20,194 & 18,683 & 15,575 & 12,383 & 11,035 & 10,075 & 13,582 & 12,642 & 11,272 & 10,666 & 9,917 & 9,295\\
    Aspect ratio & 0.59 & 0.67 & 0.67 & 0.55 & 0.64 & 0.64 & 0.84 & 0.96 & 0.90 & 0.63 & 0.70 & 0.68\\
    \bottomrule
  \end{tabular}
\end{table}

\paragraph{Metric Estimations.}
\autoref{tab:metric_results} presents results that were
measured on the final drawing of a graph. As mentioned earlier,
after the layering step these values are not available
and estimations are commonly used to deduce the
quality of a result.
For our example graphs,
the estimated area reduced from $222.9$ (\eaga)
to $187.4$ (\wwglay{1}{30}) on average.
The estimated aspect ratios increase on average from
$0.74$ to $0.84$.
Both tendencies conform to the averaged
effective values in \autoref{tab:metric_results},
\ie \acs{glpip} and the \acs{glph} perform better.
However, we observed that for 64\pc of the graphs the
tendency of the estimated area
contradicts the tendency of the effective
area.\footnote{Using \bk and \poly.}
54\pc when not considering dummy nodes.
In other words, for a specific graph
the estimated area might be decreased for \acs{glp} compared to \eaga
but the effective area is increased for \acs{glp} (or vice versa).
This clearly indicates that an estimation can be misleading.
Besides, node placement and edge routing can have a non-negligible impact
on the aspect ratio and compactness of the final drawing.

\paragraph{Performance of the Heuristic.}
Results for final drawings using the presented
heuristic are included in \autoref{tab:metric_results}
and are comparable to \wwglay{1}{30},
\ie the heuristic performs better than \eaga\
\wrt the desired metrics.

Tab.~\ref{tab:random_results} and Tab.~\ref{tab:highar_results}
underline this result and show that
the improvement step of the heuristic clearly
improves on all measured metrics.
Further, more detailed results, can be found in the appendix.
Nevertheless, the heuristic yields
significantly more reversed edges.
When aiming for compactness,
we consider this to be acceptable.

\paragraph{Execution Times.}
\label{sec:execution_times}
To solve the IP model we used CPLEX 12.6
and executed the evaluations
on a server with an Intel Xeon E5540 CPU and 24 GB memory.
The execution times for \acs{glpip} vary between 476ms for a graph
with 19 nodes and 541s for a graph with 58 nodes and
exponentially increase with the graph's node count.
This is impracticable for interactive tools that rely on automatic layout,
but is fast enough to collect optimal results for
medium sized graphs.

The execution time of the heuristic is compared to \eaga
and was measured on a laptop with an Intel i7-3537U CPU and 8 GB memory.
The reported time includes only the first two steps
of the layer-based approach.
It turns out that the execution time of the heuristic is
on average 2.3 times longer than \eaga. This
seems reasonable, as it involves two executions
of the network simplex layering method.
For the tested graphs, the construction and improvement
steps of the heuristic hardly contribute to
its overall execution time.
The effective execution time ranges between 0.1ms and
10.0ms for \eaga and 0.3ms and 19.7ms for the heuristic.
Hence, the heuristic is fast enough to be used in
interactive tools.

We also ran the algorithm five times for five randomly
generated graphs with 1000 nodes and 1500 edges.
\eaga required an average of 374ms, the heuristic 666ms with
about 4ms for construction an 2ms for improvement.
This shows that the time contribution of the
latter two is negligible even for larger graphs.

\section{Conclusion}
\label{sec:conclusion}
In this paper we address problems
with current methods for the first two phases
of the layer-based layout approach.
We argue that separately performing cycle breaking and
layering is disadvantageous when aiming for compactness.

We present a configurable method for the layering phase
that, compared to other state-of-the-art methods,
shows on average improved
performance on compactness. That is, the number of
dummy nodes is reduced significantly for most graphs
and can never increase. While the number
of dummy nodes only allows for an estimation of the
area, the effective area of the final drawing
is reduced as well.
Furthermore, graph instances for which current methods yield
unfavorable aspect ratios can easily be improved.
Also, the presented heuristic clearly improves on the
desired metrics.
Depending on the application,
a slight increase in the number of
reversed edges is often acceptable.

We want to stress that the common practice to
determine the quality of methods developed for
certain phases of the layer-based approach based
on metrics that represent estimations of the
properties of the final drawing is error-prone.
For instance, estimations of the area and aspect
ratio after the layering phase can vary significantly
from the effective values of the final drawing
and strongly depend on the used strategies
for computing node and edge coordinates.

Future work will include improving the heuristic,
\eg selecting the initial node based on a certain
criterion instead of randomly in \autoref{alg:constructLayering}
should improve the results.
We also plan to incorporate hard bounds on the width
of a drawing.
It is important that methods support
to prevent, or at least to strongly penalize,
the reversal of certain edges,
since certain diagram types
demand several edges to be drawn forwards.
Also, user studies could help understand which
edges are natural candidates to be reversed
from a human's perspective.

Furthermore, in an accompanying technical report we present
a variation of \acs{glp} where we fix the size
of the \acs{fas} while remaining free in the
choice of which edges to reverse~\cite{RueeggESvH15},
which so far has only been evaluated using integer programming.

\subsubsection{\textbf{Acknowledgements.}}
We thank Chris Mears for support regarding the initial IP formulation.
We further thank Tim Dwyer and Petra Mutzel for valuable discussions.
This work was supported by the German Research Foundation under
the project \emph{Compact Graph Drawing with Port Constraints}\\
(ComDraPor, DFG HA 4407/8-1).

\bibliographystyle{abbrv}
\bibliography{cau-rt,rts-arbeiten,pub-rts}

\pagebreak
\appendix

\section{Appendix}

\subsection{Example Drawings of North Graphs}

\begin{figure}[htb]
 \centering
 \subfloat[\textsf{g.75.1}]{
  \begin{minipage}{.4\textwidth}
    \centering
    \includegraphics[scale=0.15]{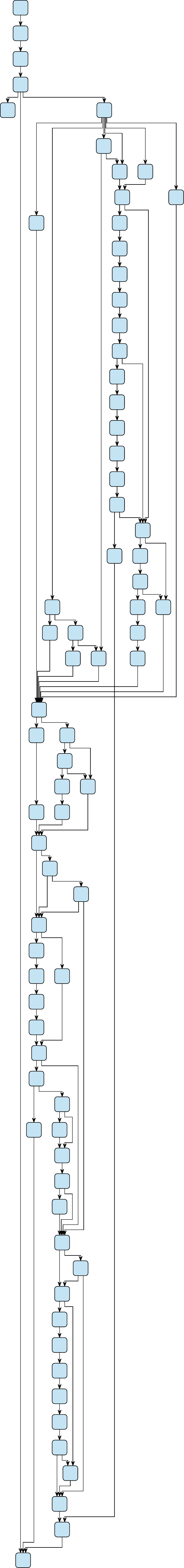}
      \hfill
    \includegraphics[scale=0.15]{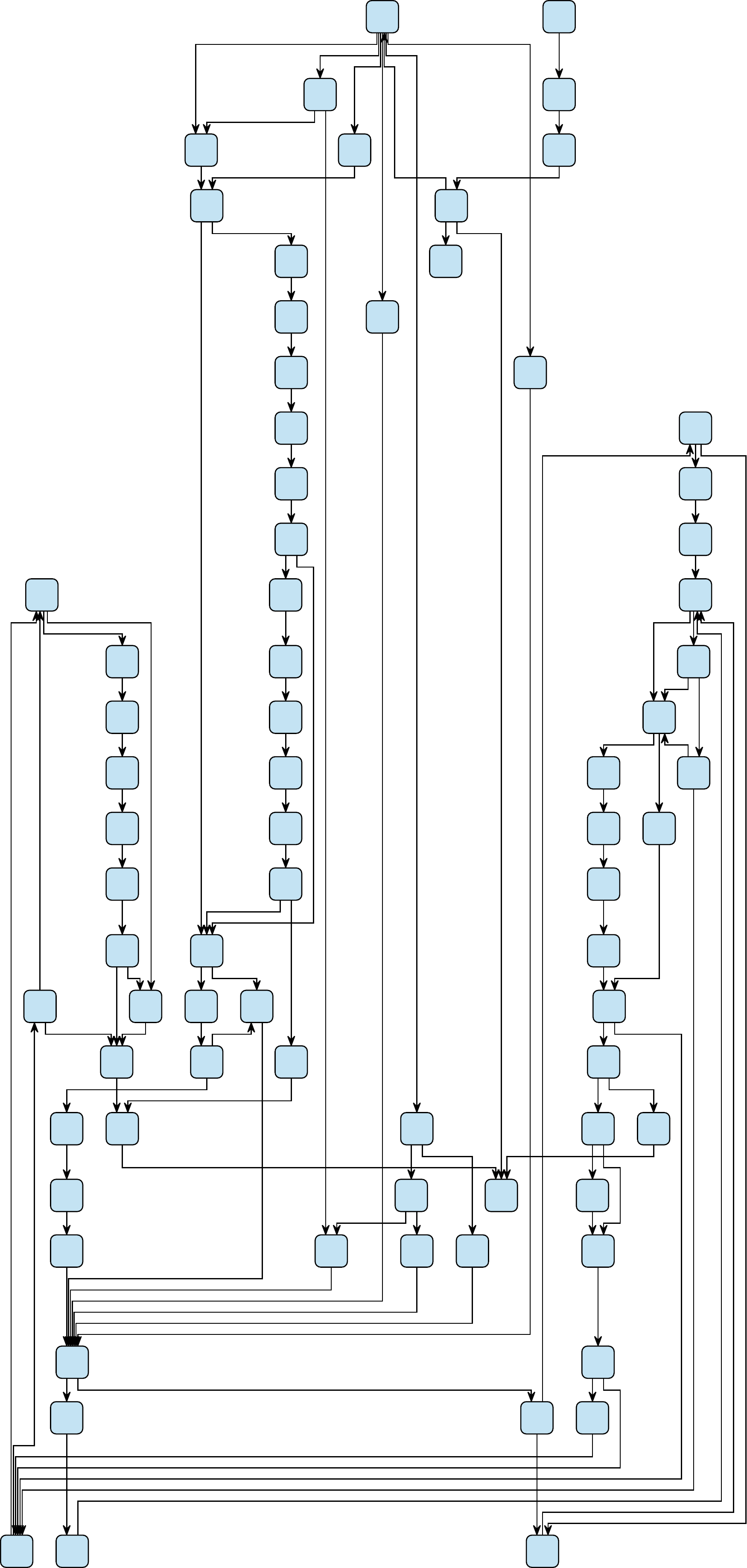}
  \end{minipage}
 }\hfill
 \subfloat[\textsf{g.80.5}]{
  \begin{minipage}{.4\textwidth}
    \centering
    \includegraphics[scale=0.15]{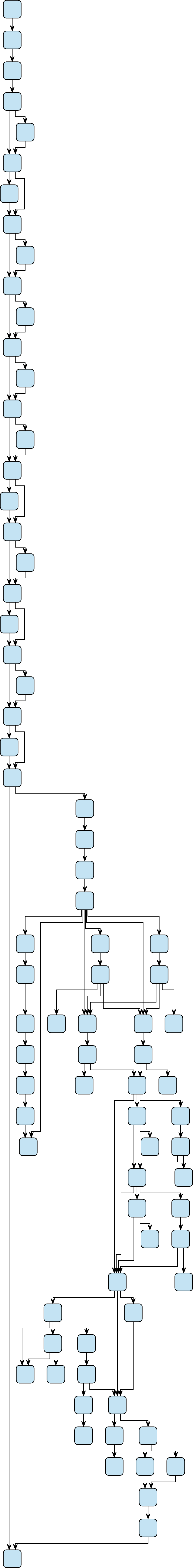}
      \hfill
    \includegraphics[scale=0.15]{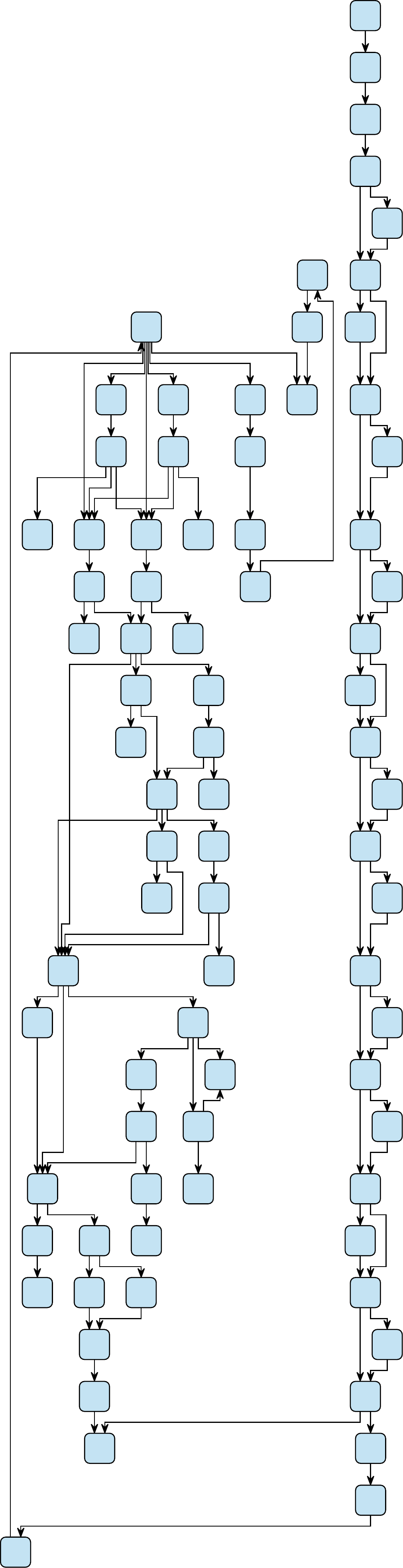}
  \end{minipage}
 }
 \caption{For each graph the left drawing is produced using
 \eaga and the right drawing using the \acs{glph} as presented here.
 Graphs are taken from the North graphs library.}
\end{figure}

\begin{figure}[htb]
 \centering
 \subfloat[\textsf{g.99.0}]{
  \begin{minipage}{.42\textwidth}
    \centering
    \includegraphics[scale=0.13]{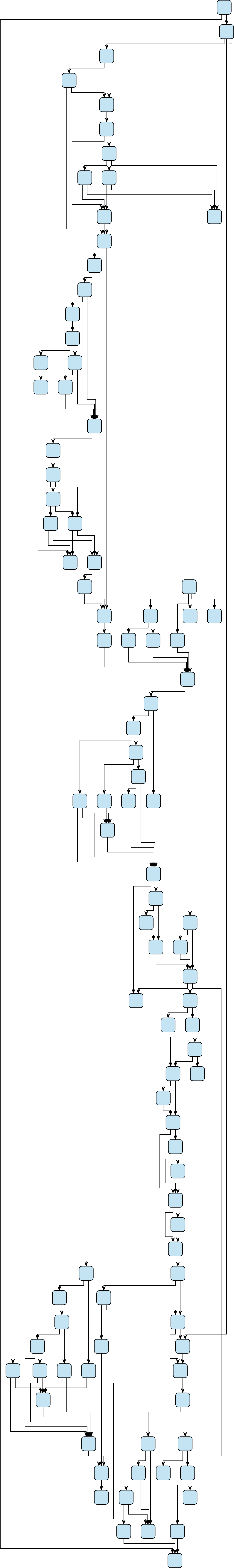}
      \hfill
    \includegraphics[scale=0.13]{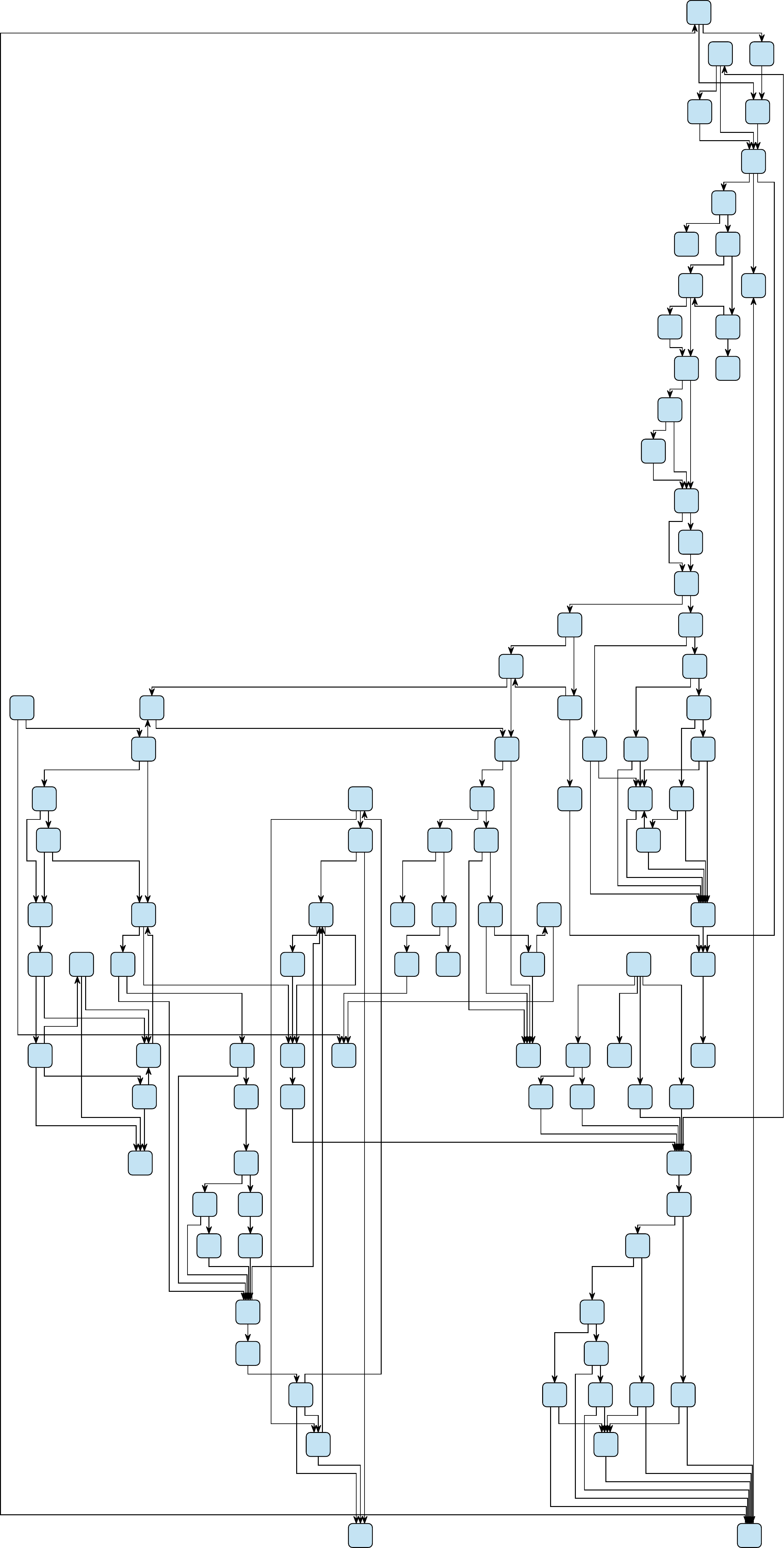}
  \end{minipage}
 }\hfill
 \subfloat[\textsf{g.86.5}]{
  \begin{minipage}{.42\textwidth}
    \centering
    \includegraphics[scale=0.13]{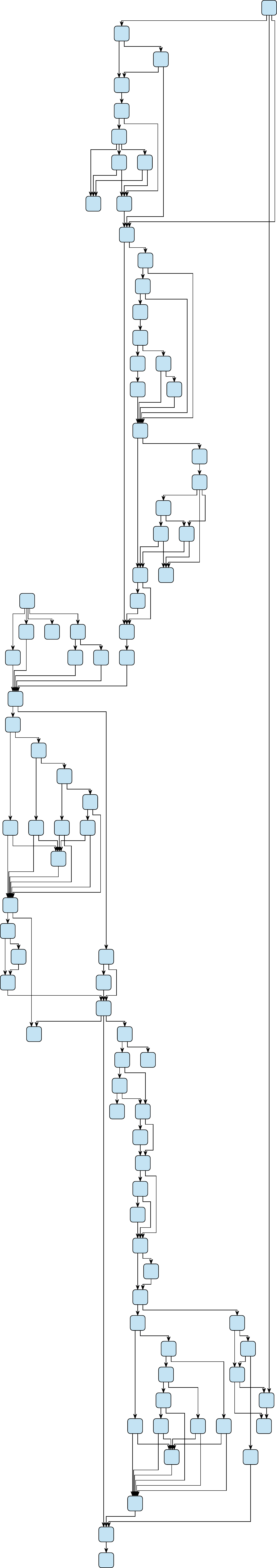}
      \hfill
    \includegraphics[scale=0.13]{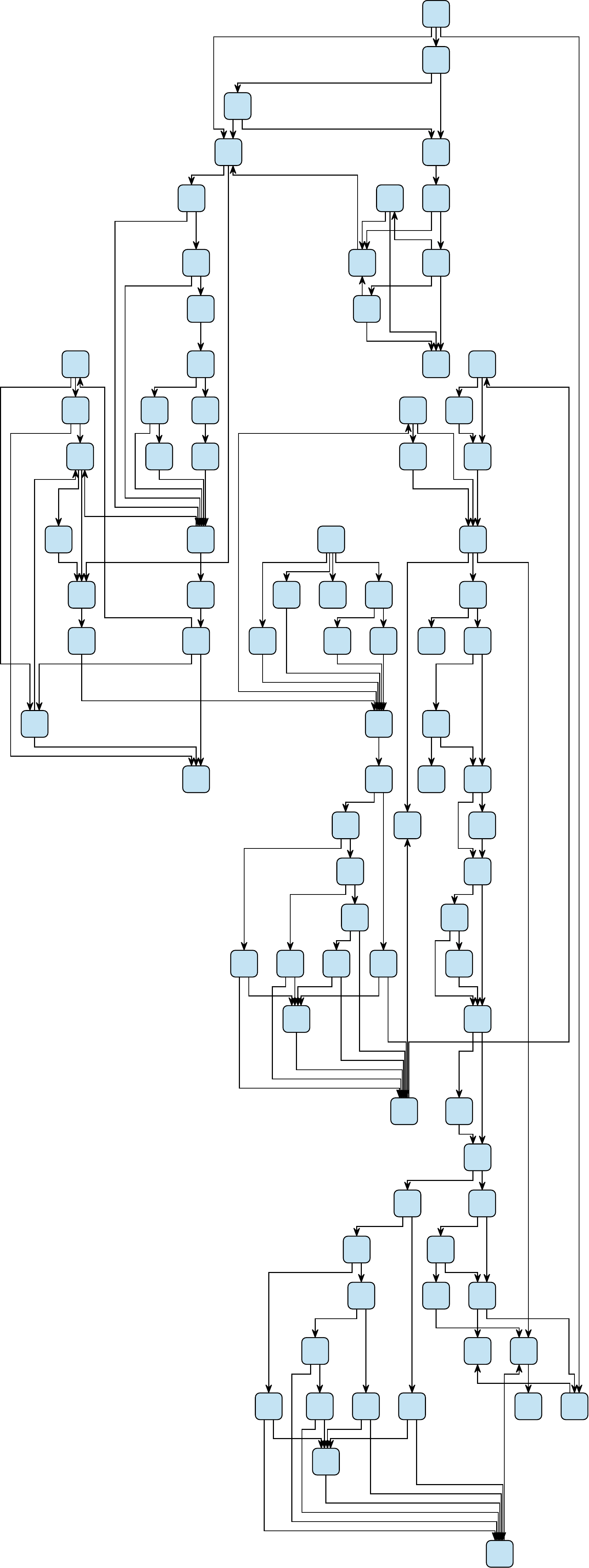}
  \end{minipage}
 }
 \caption{For each graph the left drawing is produced using
 \eaga and the right drawing using the \acs{glph} heuristic as presented here.
 Graphs are taken from the North graphs library.}
\end{figure}

\subsection{Detailed Results}
\label{sec:detailed_results}
\begin{figure}[H]
  \centering
  \scalebox{1}{
    \input{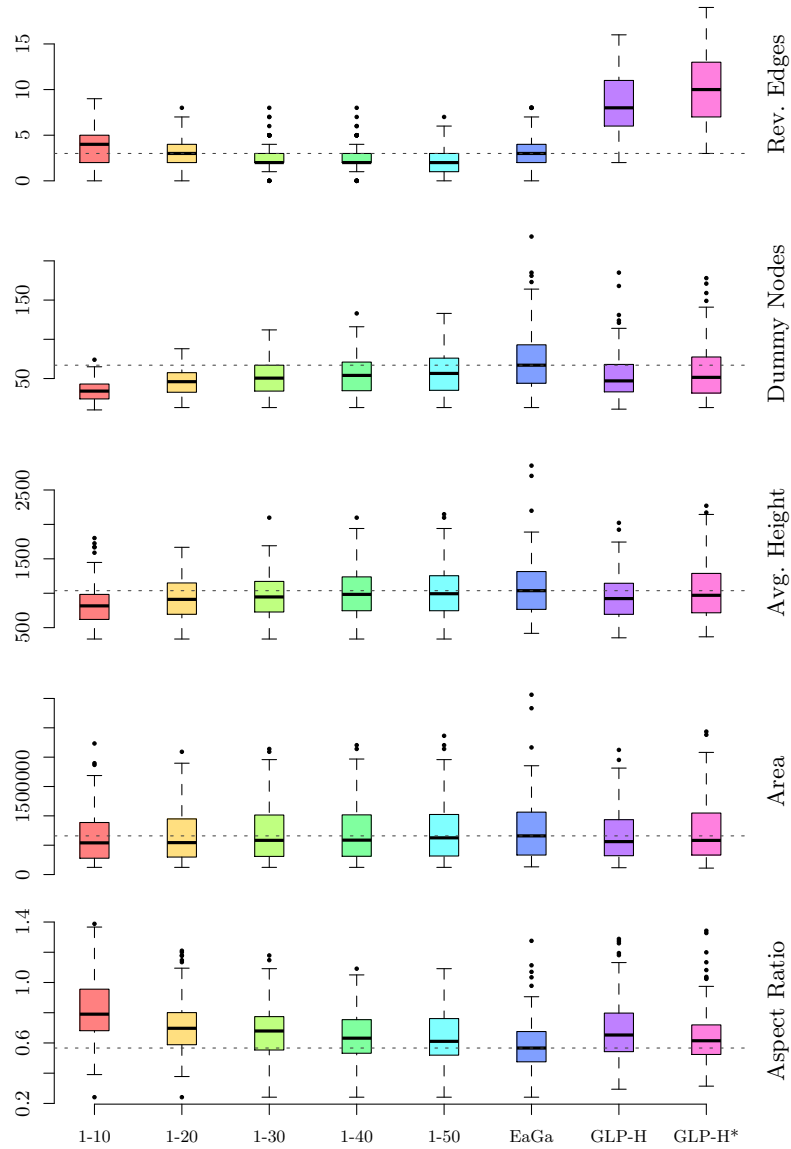}
  }
  \caption{
    \textbf{Random graphs}: Detailed results in the form of boxplots.
    A summary can be seen in Tab.\ref{tab:random_results}.
    The dashed line represents the median of \eaga.
    Lower values are better, with the exception of the aspect ratio.
    It can be seen that the methods presented here,
    improve the drawing \wrt the relevant metrics.
    It is noteworthy that for \wwglay{1}{30}, \wwglay{1}{40}, and \wwglay{1}{50},
    both the number of reversed edges and the number of dummy nodes
    is smaller compared to \eaga.
  }
  \label{fig:boxplot_random_detailed1}
\end{figure}

\begin{figure}[H]
  \centering
  \scalebox{1}{
    \input{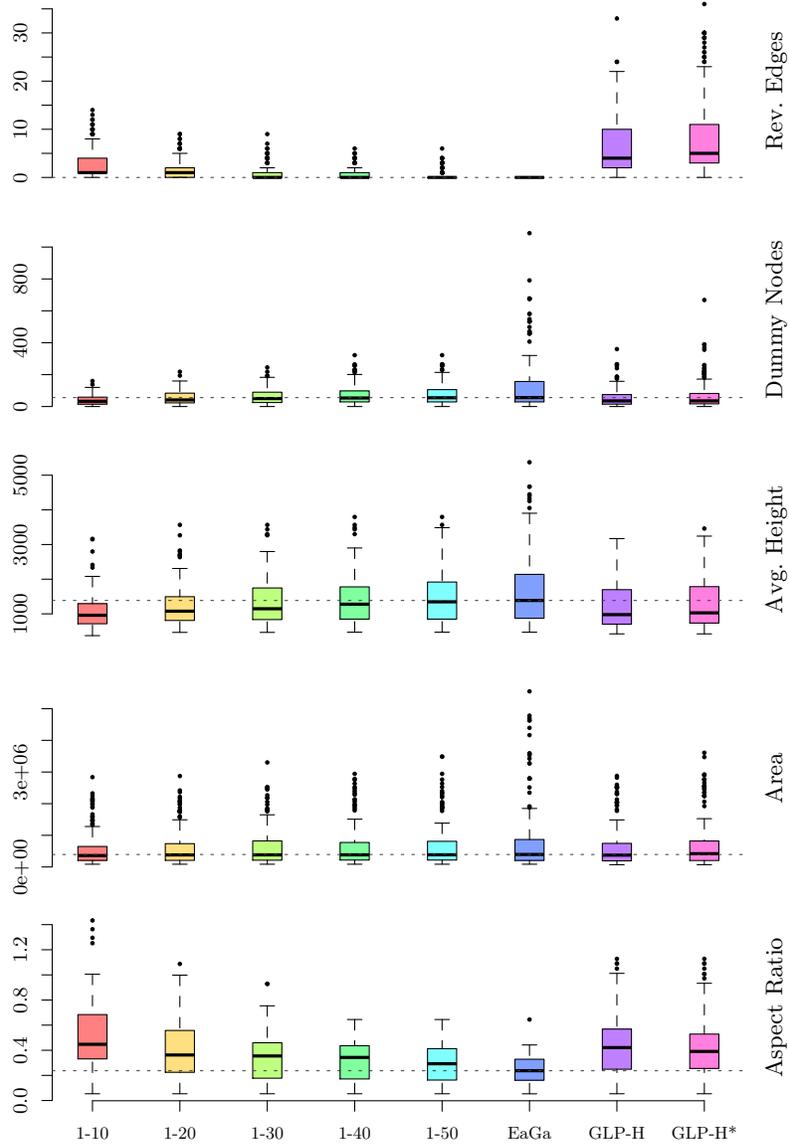}
  }
  \caption{
    \textbf{North graphs}: Detailed results in the form of boxplots.
    A summary can be seen in Tab.\ref{tab:highar_results}.
    The dashed line represents the median of \eaga.
    Lower values are better, with the exception of the aspect ratio.
    It can be seen that the methods presented here,
    improve the drawing \wrt the relevant metrics.
    The North graphs are acyclic which is why \eaga
    consistently produces zero reversed edges.
    The aspect ratio of the majority of the graphs
    improves significantly with a constant to slightly
    improved area.
  }
  \label{fig:boxplot_random_detailed2}
\end{figure}

\begin{figure}[H]
  \centering
  \scalebox{1}{
    \input{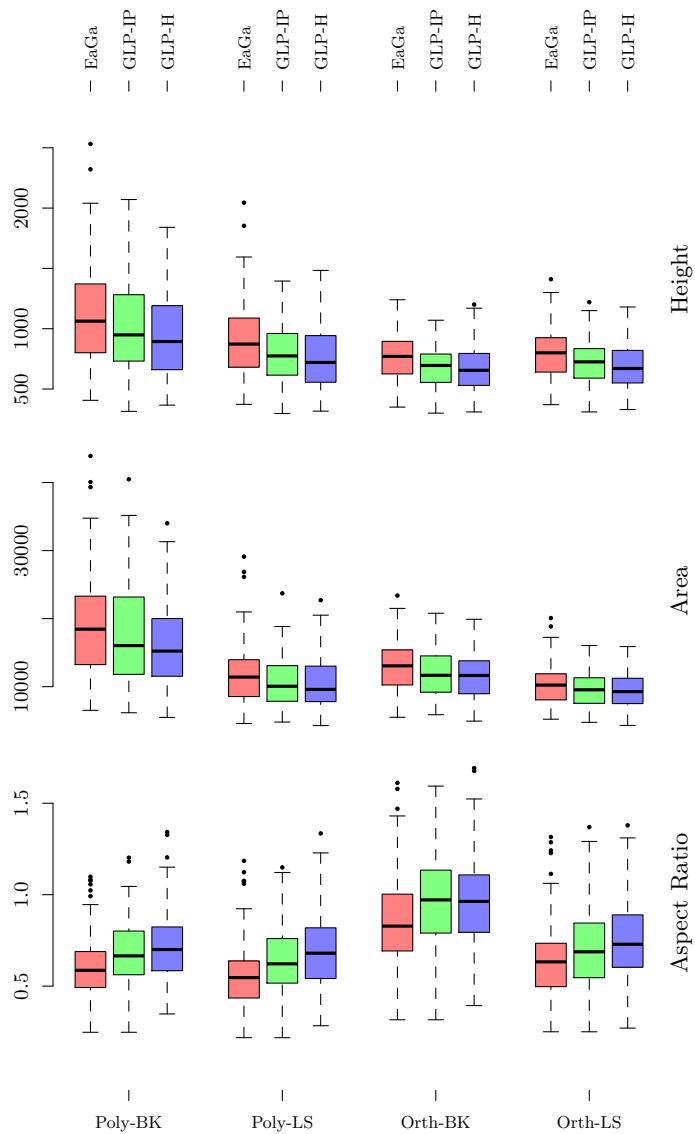}
  }
  \caption{
    Detailed results for the produced drawings using
    different strategies
    for the layer-based approach's phases
    as discussed in \autoref{sec:evaluations}
    (\cf \autoref{tab:metric_results}).
    It can be seen that for every combination
    \acs{glpip} and \acs{glph} improve \wrt
    the tested metrics when compared to \eaga.
    Furthermore, the results emphasize that
    different strategies can result in
    significantly different drawings,
    especially when it comes to aspect ratio.
    For instance, orthogonal-style edges allow
    for less height and area.
    Node coordinates assigned by LS tend to
    allow for smaller area than BK.
  }
  \label{fig:boxplot_metrics_detailed}
\end{figure}

\end{document}